\documentclass[twocolumn,superscriptaddress,aps,prl]{revtex4-1}
\usepackage{amsfonts}
\usepackage{amssymb}
\usepackage{amsmath}
\usepackage{graphicx}
\usepackage{epsfig}
\usepackage{makecell}
\usepackage{enumerate}
\usepackage{mathrsfs}

\begin{document}

\title{Antihelical Edge States in Two-dimensional Photonic Topological Metals%
}
\author{Liangcai Xie}
\author{Liang Jin}
\email{jinliang@nankai.edu.cn}
\author{Zhi Song}

\begin{abstract}
Topological edge states are the core of topological photonics. Here we
introduce the antihelical edge states of time-reversal symmetric topological
metals and propose a photonic realization in an anisotropic square lattice
of coupled ring resonators, where the clockwise and counterclockwise modes
play the role of pseudospins. The antihelical edge states robustly propagate
across the corners toward the diagonal of the square lattice: The same
(opposite) pseudospins copropagate in the same (opposite) direction on the
parallel lattice boundaries; the different pseudospins separate and converge
at the opposite corners. The antihelical edge states in the topological
metallic phase alter to the helical edge states in the topological
insulating phase under a metal-insulator phase transition. The antihelical
edge states provide a unique manner of topologically-protected robust light
transport applicable for topological purification. Our findings create new
opportunities for topological photonics and metamaterials.
\end{abstract}

\affiliation{School of Physics, Nankai University, Tianjin 300071, China}
\maketitle

\textit{Introduction.---}The fundamental concepts in condensed matter physics introduced to
topological photonics inspire the rapid development of photonic topological
states \cite{LLu14,ABKNPo17,TORMP19}. The chiral edge states of topological
insulators unidirectionally propagate along boundaries and require the
breaking of time-reversal symmetry \cite%
{FHPRL08,ZWNa09,MHNP11,KFNP12,GQLPRL13,SZhangPRL20}. The degenerate
clockwise and counterclockwise modes of ring resonators experience opposite
artificial magnetic fields and provide a pseudospin degree of freedom \cite%
{MHNP11}. The helical edge states of time-reversal symmetric topological
insulators with different pseudospins unidirectionally propagate in opposite
directions. The edge states of topological metals are
topologically-protected in the gapless phase. Antichiral edge states have
been proposed and implemented on zigzag edges by modifying the
next-nearest-neighbor hopping phase of the Haldane model \cite%
{FranzPRL18,ZhangBLPRL20,ChongYD21,LiZYPRL22}. Antichiral edge states
propagate along the parallel lattice boundaries in the same direction.
Recent progress in antichiral edge states has led to the development of a
solution for creating topological metals. Topological metals are difficult
to create because of the challenge of separating gapless bands and breaking
time-reversal symmetry in photonics. Thus, is it possible to have
time-reversal symmetric topological metals with topologically-protected edge
states and robust propagation?

\begin{figure*}[tb]
\includegraphics[bb=0 0 524 142,width=18.0cm]{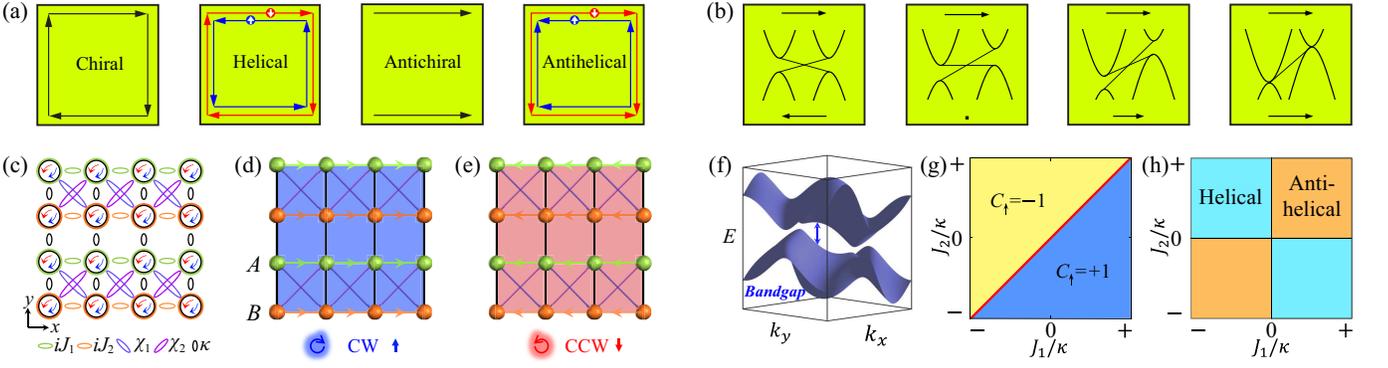}
\caption{(a) Robust two-dimensional propagation of chiral,
helical, antichiral, and antihelical edge states. Time-reversal symmetry is
broken for chiral and antichiral edge states, and time-reversal symmetry is
preserved for helical and antihelical edge states. Chiral and helical edge
states are in the insulating phases; antichiral and antihelical edge states
are in the metallic phases. Different spins of antihelical edge states
separate and converge at the opposite corners on the diagonal. (b) Schematic
of band and edge state propagation for topological insulator,
metal-insulator phase transition, topological metal, and
topological phase transition. The direction and length of the arrow
indicate the velocity of edge states. At the TPT point, the velocities in
the horizontal direction on the top and bottom boundaries are identical;
across the TPT point from $C_{\uparrow }=+1$ to $C_{\uparrow }=-1$, the
velocity on the bottom boundary exceeds the velocity on the top boundary.
(c) Schematic of two-dimensional anisotropic square lattice of coupled ring
resonators. (d) Schematic of Hamiltonian for pseudospin-up. (e) Schematic of
Hamiltonian for pseudospin-down. For $\protect\chi =\protect\kappa /2=1/2$, $%
J_{1}=1$, $J_{2}=1/2$, and $\protect\delta =3/2$, (f) nonzero $\protect\chi $
opens the band gap, (g) is the phase diagram for the pseudospin-up, and (h)
is the distribution of edge states.}
\label{fig1}
\end{figure*}

Here, we introduce the antihelical edge states [Fig.~\ref{fig1}(a)] of
time-reversal symmetric topological metals and a photonic realization is
proposed in the two-dimensional anisotropic square lattice of coupled ring
resonators. The pseudospins are time-reversal symmetric counterparts, and
the introduction of pseudospins addresses the difficulty of breaking
time-reversal symmetry in photonics. The symmetric component of
next-nearest-neighbor couplings creates a nontrivial topology, and the
anti-symmetric component of next-nearest-neighbor couplings separates energy
bands and supports the antihelical edge states on both horizontal and
vertical boundaries. Antihelical edge states robustly copropagate along
corners with opposite pseudospins that separate and converge at the opposite
corners on the diagonal of a lattice. The antihelical edge states become
helical edge states during a metal-insulator phase transition [Fig.~\ref%
{fig1}(b)]. Photonic topological metals provide a new direction for research
on topological photonics.

\textit{Anisotropic square lattice of ring resonators.---}Figure~\ref{fig1}(c) presents a two-dimensional anisotropic square lattice
of coupled ring resonators (see Supplementary materials A). The clockwise mode
(pseudospin-up) and counterclockwise mode (pseudospin-down) are
time-reversal counterparts, and they experience the opposite Peierls phases
in the horizontal couplings between the nearest-neighbor resonators \cite%
{MHNP11}. The lattice for the pseudospin-up (pseudospin-down) is presented
in Fig.~\ref{fig1}(d) [Fig.~\ref{fig1}(e)]. The horizontal couplings $\pm
iJ_{1}$ indicated in green and $\pm iJ_{2}$ indicated in orange are
tunneling-direction-dependent, break the time-reversal symmetry of the
Hamiltonian for each individual pseudospin, and separately affect the edge
states localized on the upper and lower boundaries. The vertical coupling
between the nearest-neighbor resonators is $\kappa $ indicated in black. The
cross couplings between the next-nearest-neighbor resonators are $\chi
_{1}=\chi +\delta $, $\chi _{2}=\chi -\delta $ \cite{LeykamPRL18}. The
symmetric component $\chi $ opens a band gap [Fig.~\ref{fig1}(f)] and
creates a nontrivial topology, whereas the anti-symmetric component $\delta $
affects the band structure, separates the bands in both the $x$ and $y$
directions, and creates antihelical edge states that are localized on the
left and right boundaries in the topological metallic phase.

\textit{Topological phases.---}The Bloch Hamiltonian for the pseudospin-up is $h_{\uparrow }\left( \mathbf{k%
}\right) =d_{0}\left( \mathbf{k}\right) \sigma _{0}+\mathbf{d\left( \mathbf{k%
}\right) \cdot \sigma }$, where $\sigma _{0}$ is the identical matrix and $%
\mathbf{\sigma =}\left( \sigma _{x},\sigma _{y},\sigma _{z}\right) $ is the
Pauli matrix. The first term adjusts the band energy and the second term
determines the band topology. We have $d_{0}\mathbf{\left( \mathbf{k}\right) 
}=(J_{1}+J_{2})\sin k_{x}$ and $\mathbf{d}\left( \mathbf{k}\right) =\mathbf{r%
}_{1}(k_{x})-\mathbf{r}_{2}(k_{y})$ with $\mathbf{r}_{1}(k_{x})=(\kappa
+2\chi \cos k_{x},-2\delta \sin k_{x},(J_{1}-J_{2})\sin k_{x})$, $\mathbf{r}%
_{2}(k_{y})=(-\kappa \cos k_{y},\kappa \sin k_{y},0)$. $h_{\uparrow }\left( 
\mathbf{k}\right) $ respects the particle-hole symmetry $\sigma
_{z}h_{\uparrow }^{T}\left( \mathbf{k}\right) \sigma _{z}^{-1}=-h_{\uparrow
}\left( -\mathbf{k}\right) $. The topological phase belongs to the class $D$
and it is characterized by a $\mathbb{Z}$ topological invariant, that is,
the spin-Chern number $C_{\uparrow }=-\left( 4\pi \right) ^{-1}\int
\left\vert \mathbf{d}\right\vert ^{-3}\mathbf{d}\cdot \left( \partial
_{k_{x}}\mathbf{d}\times \partial _{k_{y}}\mathbf{d}\right) dk_{x}dk_{y}$
[Fig.~\ref{fig1}(g)]. The band topology is captured by the effective
magnetic field $\mathbf{d}\left( \mathbf{k}\right) $. After $\mathbf{d}%
\left( \mathbf{k}\right)$ is substituted with $\mathbf{r}_{1}(k_{x})-\mathbf{%
r}_{2}(k_{y})$ in $C_{\uparrow }$, the spin-Chern number becomes the
definition of a linking number of the two independent periodic vectors $%
\mathbf{r}_{1}(k_{x})$ and $\mathbf{r}_{2}(k_{y})$ (see Supplementary materials B). $%
\mathbf{r}_{1}(k_{x})$ is an ellipse that passes through $\left( \kappa
-2\chi ,0,0\right) $ and $\left( \kappa +2\chi ,0,0\right) $. $\mathbf{r}%
_{2}(k_{y})$ is a circle centered at the origin with a fixed radius $%
\left\vert \kappa \right\vert $ in the $z=0$ plane. Thus, the two closed
curves $\mathbf{r}_{1}(k_{x})$ and $\mathbf{r}_{2}(k_{y})$ are linked when $%
0<\left\vert \chi \right\vert <\left\vert \kappa \right\vert $.

\begin{figure*}[tb]
\includegraphics[bb=7 0 580 185,width=18.0cm]{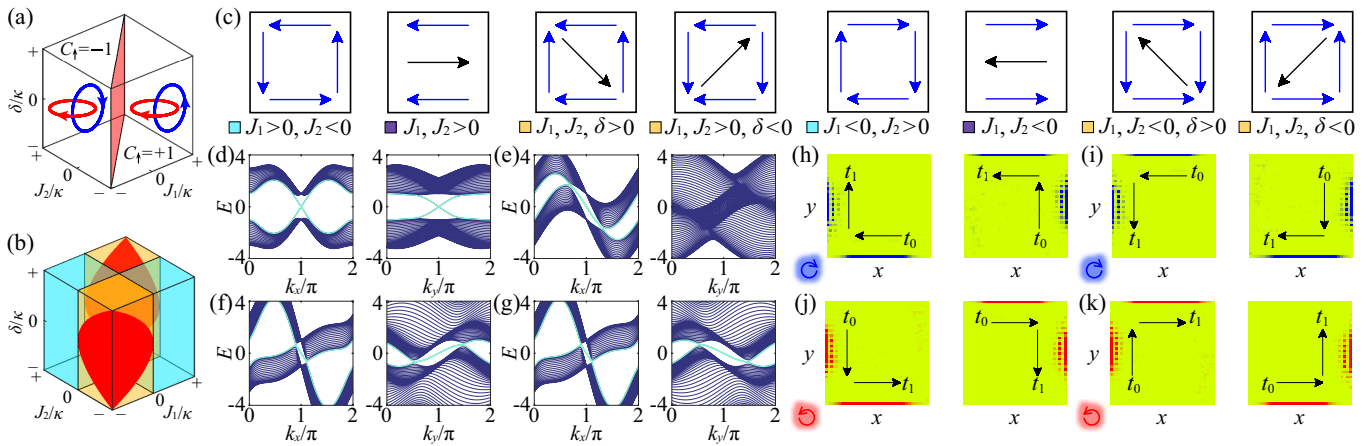}
\caption{(a) Spin-Chern numbers of topological phases
indicated by nontrivial topologies of links. Topological phase transition
occurs at the red plane. (b) Types of edge states. The entire parameter
space $J_{1}$-$J_{2}$-$\protect\delta $ is divided into three regions. The
cyan regions $J_{1}J_{2}<0$ are the insulating phase, the red regions $%
\protect\delta ^{2}<J_{1}J_{2}$ are the metallic phase with the separable
bands being present only in the horizontal direction, and the orange regions 
$0<J_{1}J_{2}<\protect\delta ^{2}$ are the metallic phase with the separable
bands being present in both directions. (c) Propagation of antihelical and
helical edge states for pseudospin-up excitation; the black arrows indicate
the propagation in the bulk, and they pertain to the topological phase $0<%
\protect\chi <\protect\kappa $. (d)-(g) Open boundary spectra of the
pseudospin-up for the four cases on the left-half of (c). (d) $J_{2}=-1,%
\protect\delta =1/2$, (e) $J_{2}=1,\protect\delta =1/2$, (f) $J_{2}=1,%
\protect\delta =3/2$, (g) $J_{2}=1,\protect\delta =-3/2$. The other
parameters are $\protect\kappa =1$, $J_{1}=1$, and $\protect\chi =1/2$. The
simultaneous mirror reflection of the band structures with respect to $%
k_{x}=0$ and $k_{y}=0$ in (d)-(g) are the four cases on the right-half of
(c). (h)-(k) Robust transports of antihelical edge states performed in a $%
40\times 40$ square lattice for (f) and (g) at the interface between two
topological phases $C_{\uparrow}=\pm 1$, where the velocities along the top
and bottom boundaries in the horizontal direction are the same. Dynamics are
robust to random coupling disorder (see Supplementary materials C). (h) and (i)
pertain to pseudospin-up excitation; (j) and (k) pertain to pseudospin-down
excitation. (h) and (j) are for case (f), (i) and (k) are for case (g).}
\label{fig2}
\end{figure*}

In Fig.~\ref{fig2}(a), the red circle $\mathbf{r}_{2}(k_{y})$ that is fixed
on the $z=0$ plane is always clockwise; the rotation direction of the blue
ellipse $\mathbf{r}_{1}(k_{x})$ reverses along the topological phase
transition plane $J_{1}=J_{2}$, where the band gap vanishes and $\mathbf{r}%
_{1}(k_{x})$ becomes coplanar to $\mathbf{r}_{2}(k_{y})$. The blue ellipse $%
\mathbf{r}_{1}(k_{x})$ circles counterclockwise around the red circle $%
\mathbf{r}_{2}(k_{y})$ once in the region $J_{1}>J_{2}$ ($C_{\uparrow }=+1$%
); and the blue ellipse $\mathbf{r}_{1}(k_{x})$ circles clockwise around the
red circle $\mathbf{r}_{2}(k_{y})$ once in the region $J_{1}<J_{2}$ ($%
C_{\uparrow }=-1$). The right-hand (left-hand) rule identifies $C_{\uparrow
}=+1$ ($C_{\uparrow }=-1$): orient the thumb pointing along the arrow of the
red circle, curl the rest of the fingers pointing along the arrow of the
blue ellipse. Notably, in the phase diagram [Fig.~\ref{fig2}(b)], two
topological phases with opposite spin-Chern numbers have an identical number
of edge states.

\textit{Antihelical edge states.---}Topological edge states appear in the situation that the
next-nearest-neighbor coupling $\chi $ is weak than the nearest-neighbor
coupling $\kappa $. The nontrivial topology supports helical and antihelical
edge states [Fig.~\ref{fig1}(h)] that are different from their band
structures and their ways of propagation. The antihelical (helical) edge
states on the parallel boundaries propagate in the same (opposite)
direction, being robust against disorder because of the topological
protection and the spatial separation between the edge and bulk states \cite%
{FranzPRL18}.

In Fig.~\ref{fig1}(b), the regions $J_{1}J_{2}<0$ are the topological
insulating phase; the helical edge states with different pseudospins
propagate clockwise or counterclockwise along the lattice boundaries. The
topological phase undergoes a metal-insulator phase transition at $%
J_{1}J_{2}=0$. The regions $J_{1}J_{2}>0$ are the topological metallic phase 
\cite{KamenevPRL18}; the edge states excited by the pseudospin-up propagate
toward the left (or right) on the two parallel boundaries, but the edge
states excited by the pseudospin-down propagate toward the right (or left)
boundaries for $J_{1},J_{2}>0$ (or $J_{1},J_{2}<0$). These edge states
associated with the two pseudospins constitute the antihelical edge states,
and the required counterpropagating modes for the antihelical edge states
are the bulk states. The bands touch at the topological phase transition,
where topological edge states are degenerate and exhibit identical
dispersion and propagation velocity.

The square lattice exhibits nontrivial topology throughout the parameter
space $J_{1}$-$J_{2}$-$\delta $ [Fig.~\ref{fig2}(a)]. The various types of
edge states, as distinguished by their propagation, are presented in Fig.~%
\ref{fig2}(b). The helical edge states are in $J_{1}J_{2}<0$ indicated in
cyan and the antihelical edge states are in the other regions $J_{1}J_{2}>0$%
. The antihelical edge states differ from the helical edge states in term of
their copropagation on the parallel boundaries \cite{MHNP11,FranzPRL18}. The
antihelical edge state excitations with different pseudospins
unidirectionally propagate in the opposite directions along the lattice
boundaries and enable the separation of the robustly copropagated chiral
mode. The helical edge states in the topological insulating phase always
appear in both the horizontal and vertical directions; however, the
antihelical edge states in the topological metallic phase usually appear
only in one direction because of the inseparable band energy in the other
direction. Here, the antihelical edge states are simultaneously present in
both the horizontal and vertical directions for $\delta ^{2}>J_{1}J_{2}>0$,
where the energy bands are separable in both directions. The antihelical
edge states are only present in the horizontal directions for $\delta
^{2}<J_{1}J_{2}$. In Fig.~\ref{fig2}(b), the surfaces $J_{1}J_{2}=0$ and $%
\delta ^{2}=J_{1}J_{2}$ divide the parameter space $J_{1}$-$J_{2}$-$\delta $
into three phases with eight regions. The eight types of edge state
propagation for the pseudospin-up excitations are presented in Fig.~\ref%
{fig2}(c). The pseudospin-down edge state excitations propagate in the
opposite directions.

The simultaneous sign changes of $J_{1}$ and $J_{2}$ reverse the propagation
directions of all the antihelical edge states. The four cases of edge modes
on the left-half and right-half of Fig.~\ref{fig2}(c) propagate in opposite
directions. The sign change of $\delta $ only alters the propagation
direction of the antihelical edge states in the vertical direction. The
antihelical edge states at the strong $\chi _{1}$ ($\delta >0$) pass along
the bottom-left and top-right corners and scatter into the bulk at the
top-left and bottom-right corners; by contrast, the antihelical edge states
at the strong $\chi _{2}$ ($\delta <0$) pass along the bottom-right and
top-left corners and scatter into the bulk at the top-right and bottom-left
corners. In Fig.~\ref{fig2}(c), the helical edge states propagate
counterclockwise (clockwise) along the lattice boundaries for $C_{\uparrow
}=+1$ ($C_{\uparrow }=-1$) in the cyan region $J_{1}>0,J_{2}<0$ ($%
J_{1}<0,J_{2}>0$). The antihelical edge states along the horizontal
direction in the red region $J_{1},J_{2}>0$ ($J_{1},J_{2}<0$) of $\delta
^{2}<J_{1}J_{2}$; the edge state excitation propagates leftward (rightward),
scatters into the bulk at the corners, and goes backward. The antihelical
edge states are present along both the horizontal and vertical directions
when $0<J_{1}J_{2}<\delta ^{2}$; the edge state excitation propagates half a
closed-loop along the lattice boundaries, scatters into the bulk at the
corners, and goes backward along the diagonal direction with the support of
the scattering states. The four cases are respectively distributed in the
orange regions $\left( J_{1},J_{2},\delta >0\right) $, $\left(
J_{1},J_{2}<0,\delta >0\right) $, $\left( J_{1},J_{2}>0,\delta <0\right) $,
and $\left( J_{1},J_{2},\delta <0\right) $ of $0<J_{1}J_{2}<\delta ^{2}$.

Figures~\ref{fig2}(d)-\ref{fig2}(g) present the four cases of the band
structures for the pseudospin-up presented on the left-half of Fig.~\ref%
{fig2}(c). The propagation of antihelical edge states in Figs.~\ref{fig2}(f)
and~\ref{fig2}(g) for the pseudospin-up are simulated in Figs.~\ref{fig2}(h)
and~\ref{fig2}(i), and the corresponding propagation for the pseudospin-down
are simulated in Figs.~\ref{fig2}(j) and~\ref{fig2}(k). The separation and
convergence of different pseudospins toward the opposite corners of the
square lattice are observed.

In conclusion, we have introduced the antihelical edge states in the
time-reversal symmetric topological metallic phase, and proposed the
photonic realization in coupled ring resonators. The antihelical edge states
with different pseudospins propagate in the opposite directions along the
corners toward the diagonal of the square lattice. Unconventional
copropagation is applicable for the robust spin purification. The
antihelical edge states become the helical edge states after undergoing a
metal-insulator phase transition. The wide range of reconfigurable robust
light propagation enables the flexible control of light flow at the edges.
Our findings provide insight into the photonic topological metals and are
applicable for the acoustic lattices and other two-dimensional
metamaterials. Concepts pertaining to topological metals and antihelical
edge states are inspiring in the condensed matters and topological materials.

\section*{Acknowledgments}

This work was supported by National Natural Science Foundation of China
(Grants No.~12222504, No.~11975128, and No.~11874225).

\section*{Supplementary Materials}

\subsection{A. Experimental realization of the square lattice}

\begin{figure}[tb]
\includegraphics[bb=0 0 490 200,width=8.8 cm]{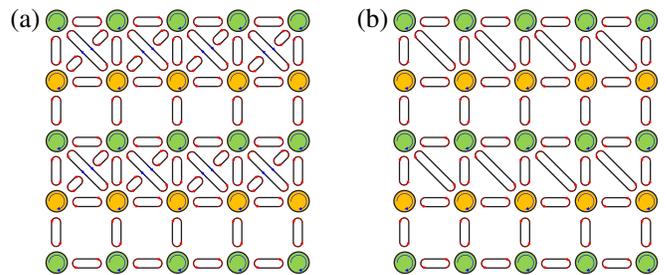}
\caption{Schematic of the 2D square lattice of coupled ring resonators. The
ring resonators in green (orange) are the sites $A$ ($B$), and the other
stadium resonators are the linking resonators. (a) $\protect\chi_1\protect%
\chi_2\neq0$. (b) $\protect\chi_2=0$. The arrows in the ring resonators
indicate the CW mode.}
\label{figS1}
\end{figure}

In this section, the experimental realization of the square lattice in the
two-dimensional (2D) coupled resonator array is discussed. Figure \ref{figS1}%
(a) is the schematic of the 2D coupled resonator array. The ring resonators
are the primary resonators for the sites of the square lattice. The
resonators in green stand for the sites $A$ and the resonators in orange
stand for the sites $B$. The linking resonators mediate the photons
tunneling among the primary resonators and induce the effective couplings
between the nearest-neighbor primary resonators and the
next-nearest-neighbor primary resonators. The linking resonators and the
primary resonators are coupled through their evanescent fields and the
coupling strengths depend on the positions between the linking resonators
and the primary resonators. For example, the couplings for the CW mode
(pseudospin-up) of the primary resonators are mediated by the CCW/CW mode of
the linking resonators.

The vertical coupling $\kappa $ is reciprocal, the path lengths for the
photons tunneling upward and downward between the neighbor ring resonators
are equal. The coupling strength is approximately characterized by $\kappa
=\kappa _{l}^{2}/\Delta _{l}$~\cite{LYouPRA17}, where $\kappa _{l}$ and $%
\Delta _{l}=\omega _{c}-\omega _{link}$ are the hopping and detuning between
the primary resonators and the linking resonators; and the frequency of the
primary (linking) resonators is $\omega _{c}$ ($\omega _{link}$).

The horizontal couplings $iJ_{1}$ and $iJ_{2}$ are nonreciprocal, carrying
the Peierls phase $e^{i\pi /2}$ in the couplings. The Peierls phase is
implemented through the optical path length difference for the photons
tunneling rightward and leftward between the neighbor ring resonators. The
CW mode photons tunneling rightward from the lower half of the linking
resonator experience an additional path length $l=\lambda /2$ than the CW
mode photons tunneling leftward from the upper half of the linking resonator 
\cite{MHNP11}, where the wave length is $\lambda $. Therefore, the
photons tunneling rightward acquire an extra phase factor $e^{i\phi
}=e^{i\pi l/\lambda }=i$ and the photons tunneling leftward acquire an extra
phase factor $e^{-i\phi }=e^{-i\pi l/\lambda }=-i$ in the front of the
horizontal couplings $iJ_{1}$ and $iJ_{2}$.

The cross coupling $\chi _{1}$ ($\chi _{2}$) between the primary resonators
on the diagonal of the square plaquette is directly (indirectly) mediated by
the CCW (CW) mode of the linking resonator along the main diagonal of the
square plaquette as indicated by the blue (red) arrows. The main diagonal of
the square plaquette refers to the line along the upper-left and the
lower-right corners of the square plaquette. The cross couplings $\chi _{1}$
and $\chi _{2}$ are independently mediated through the linking resonators
along the diagonals of the square plaquette{\ \cite{LeykamPRL18}}.

The antihelical edge states can appear in the topological metal phase of the
square lattice at the specific case $\chi _{1}\chi _{2}=0$. Thus, the single
cross coupling case is adequate for the observation of antihelical edge
states in experiments. A concrete system is $\chi =\delta $. In this
situation, one of the two cross couplings $\chi _{2}=0$ vanishes as
schematically illustrated in Fig. \ref{figS1}(b). This simplifies the setup
in the experiment and facilitates the realization of antihelical edge
states. The robust propagation with strong $\chi _{2}$ can be observed in
the situation $\chi _{1}=0$ through switching the orientation of the linking
resonator to alter the connection between the nearest-neighbor resonators on
the diagonals of the square plaquette. In this manner, the proposed square
lattice can alter between two simple situations $\chi _{2}=0$ and $\chi
_{1}=0$.

For example, we consider a simple configuration in experiment that the
resonators are neatly arranged in both horizonal and vertical directions of
the 2D square lattice. Consequently, the horizontal couplings have equal
strengths $J_{1}=J_{2}$. A candidate platform for the possible realization
of the photonic topological metal and the antihelical edge states can be
chosen as follows. In a situation $\kappa =1$, $\chi =\delta =1/2$, $%
J_{1}=J_{2}=1/4$, the two cross couplings are $\chi _{1}=1$, $\chi _{2}=0$
and the cross coupling strength equals to the vertical coupling strength $%
\chi _{1}=\kappa $. The round trip length of the resonator is about $70$ $%
\mathrm{\mu }$\textrm{m}. The resonator supports a single mode transverse
electric field at the telecom wavelength $1.55$ $\mathrm{\mu }$\textrm{m}~%
\cite{HafeziNP13}. The coupling strengths evanescently decay as the width of
the air gap between the neighbor resonators; and the coupling strengths
approximately decay from $\sim 30$ $\mathrm{GHz}$ to $\sim 5$ $\mathrm{GHz}$
for the air gap width increasing from $150$ $\mathrm{nm}$ to $250$ $\mathrm{%
nm}$~\cite{HafeZiNP19}. The topologically robust transport can be
experimentally implemented in a $20\times 20$ size square lattice of coupled
resonators at the vertical coupling strength $\kappa \sim 20$ $\mathrm{GHz}$%
, the cross coupling strength $\chi _{1}\sim 20$ $\mathrm{GHz}$, $\chi
_{2}=0 $, and the horizontal coupling strengths $J_{1}\sim 5$ $\mathrm{GHz}$%
, $J_{2}\sim 5$ $\mathrm{GHz}$. In this situation, the 2D square lattice has
a nontrivial topology because of $\chi =(\chi _{1}+\chi _{2})/2<\kappa $;
and the system is a topological metal hosting the antihelical edge states in
both horizontal and vertical directions because of $\delta ^{2}>J_{1}J_{2}>0$%
. 
\begin{figure}[tb]
\includegraphics[bb=7 0 539 230,width=8.8 cm]{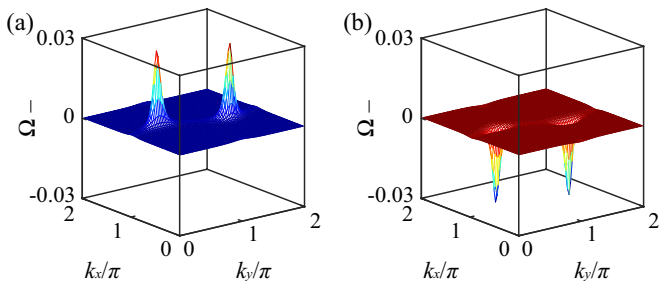}
\caption{Berry curvature in the BZ for $h_{\uparrow }\left( \mathbf{k}
\right) $. The spin-Chern number $C_{\uparrow }$ as a summation of the Berry
curvature in the entire BZ yields $C_{\uparrow }=+1$ for (a) $%
J_{1}=1,J_{2}=1/2$, and $C_{\uparrow }=-1$ for (b) $J_{1}=1,J_{2}=3/2$.
Other parameters are $\protect\kappa =1,\protect\chi =1/2$, and $\protect%
\delta =1$.}
\label{figS2}
\end{figure}

\subsection{B. Spin-Chern number characterized by the linking number of the effective magnetic field}

\begin{figure}[tb]
\includegraphics[bb=10 0 540 282, width=8.8 cm]{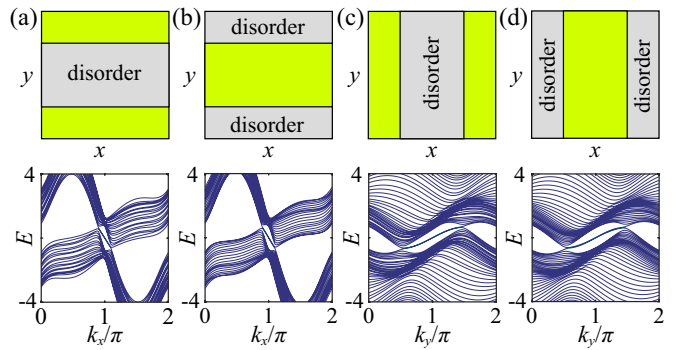}
\caption{Energy bands for random disorder in all the couplings inside and
outside the central one-half area of the square lattice. The upper panels
indicate the locations of disorder shaded in gray, and the lower panels are
the energy bands for the 1D projection lattices. (a) and (c) are for the
disorder in the bulk. (b) and (d) are for the disorder on the boundaries.
All the couplings with random disorder are deviated from the set parameters
within the range of $[-20\%,20\%]$. The parameters are $\protect\chi = 
\protect\kappa /2=1/2$, $J_{1}=J_{2}=1$, and $\protect\delta =3/2$. The
square lattice size is $40\times 40$.}
\label{figS4}
\end{figure}
\begin{figure*}[htbp]
\includegraphics[bb=10 5 595 190,width=18 cm]{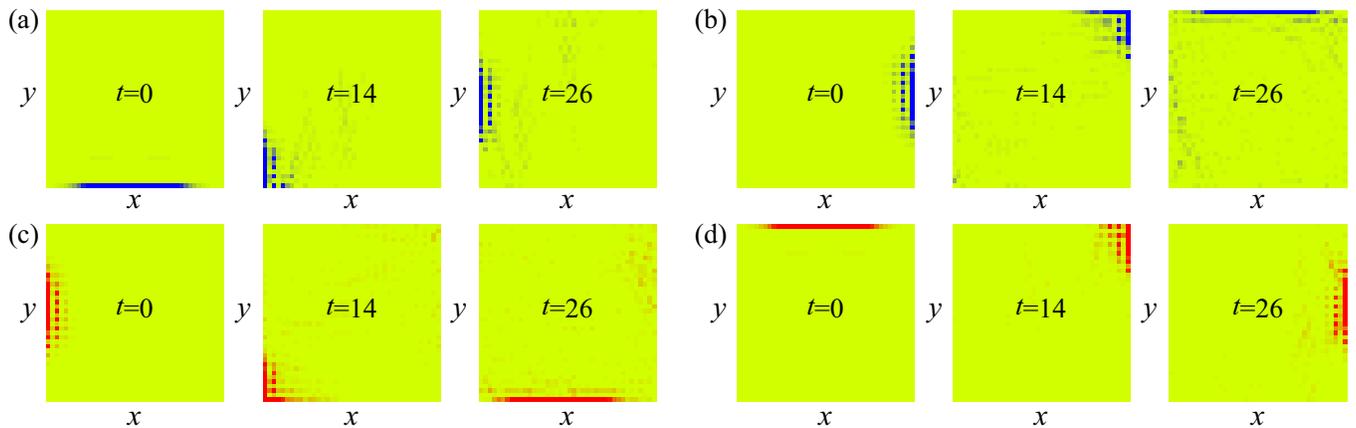}
\caption{Numerical simulations for the robust propagation in the square
lattice for random disorder in all the couplings deviated from the set
parameters within the range of $[-10\%,10\%]$. The parameters are $\protect%
\chi =\protect\kappa /2=1/2$, $J_{1}=J_{2}=1$, and $\protect\delta =3/2$.
(a) and (b) are the robust propagations of antihelical edge states for the
pseudospin-up in contrast to the robust propagations in the main text
Fig.~2(h); (c) and (d) are the robust propagations of antihelical edge
states for the pseudospin-down in contrast to the robust propagations in the
main text Fig.~2(j). The square lattice size is $40\times 40$. The unit of
time $t$ is $1/\protect\kappa$.}
\label{figS5}
\end{figure*}

The Chern number characterizes the band topology of the 2D topological phase
and the Chern number is proportional to the Hall conductance. In a 2D
time-reversal invariant system, although the total Chern number is zero, the
spin-Chern numbers are quantized. The spin-Chern number of the lower band
for the pseudospin-up (CW) mode is defined by 
\begin{equation}
C_{\uparrow }=\frac{1}{2\pi }\iint_{\mathrm{BZ}}\Omega _{-}dk_{x}dk_{y},
\label{C}
\end{equation}
where $\Omega _{-}=\nabla _{\mathbf{k}}\times \mathcal{A}_{_{\mathbf{k}}}$
is the Berry curvature, $\mathcal{A}_{_{\mathbf{k}}}=-i\left\langle \Psi
_{-}(\mathbf{k})\right\vert \nabla \left\vert \Psi _{-}(\mathbf{k}
)\right\rangle $ is the Berry connection, and $\left\vert \Psi _{-}(\mathbf{%
\ k })\right\rangle $ is the eigenstate of the lower band \cite{WGa15}. The
spin-Chern number is an integral of the Berry curvature in the entire
Brillouin zone (BZ). Figure~\ref{figS2} provides the numerical results of
the Berry curvature in the BZ for the topological phases $C_{\uparrow }=+1$
and $C_{\uparrow }=-1$ \cite{TFuku05}.

In the following, we explain the relation between the spin-Chern number and
the linking number. The spin-Chern number for the two-band Hamiltonian $%
h_{\uparrow }\left( \mathbf{k}\right) =d_{0}\sigma _{0}+\mathbf{d(k)\cdot
\sigma }$ for the pseudospin-up is expressed in the form of \cite{QWZPRB06} 
\begin{equation}
C_{\uparrow }=-\frac{1}{4\pi }\int \frac{\mathbf{d}\cdot \left( \partial
_{k_{x}}\mathbf{d}\times \partial _{k_{y}}\mathbf{d}\right) }{\left\vert 
\mathbf{d}\right\vert ^{3}}dk_{x}dk_{y}.  \label{C1}
\end{equation}%
The effective magnetic field in $h_{\uparrow }\left( \mathbf{k}\right) $ is $%
\mathbf{d(k)}=\mathbf{r}_{1}(k_{x})-\mathbf{r}_{2}(k_{y})$ with $\mathbf{r}%
_{1}(k_{x})=(\kappa +2\chi \cos k_{x},-2\delta \sin k_{x},(J_{1}-J_{2})\sin
k_{x})$ and $\mathbf{r}_{2}(k_{y})=(-\kappa \cos k_{y},\kappa \sin k_{y},0)$%
; and $d_{0}=(J_{1}+J_{2})\sin k_{x}$. Substituting $\mathbf{d(k)}=\mathbf{r}%
_{1}(k_{x})-\mathbf{r}_{2}(k_{y})$ into equation~(\ref{C1}), the spin-Chern
number is rewritten in the form of 
\begin{equation}
C_{\uparrow }=\frac{1}{4\pi }\int\nolimits_{0}^{2\pi
}\int\nolimits_{0}^{2\pi }\frac{\mathbf{r}_{1}-\mathbf{r}_{2}}{\left\vert 
\mathbf{r}_{1}-\mathbf{r}_{2}\right\vert ^{3}}\cdot \left( \frac{\partial 
\mathbf{r}_{1}}{\partial k_{x}}\times \frac{\partial \mathbf{r}_{2}}{%
\partial k_{y}}\right) dk_{x}dk_{y}.  \label{C2}
\end{equation}

In geometry, equation~(\ref{C2}) is the definition of the linking number of
two independent closed curves $\mathbf{r}_{1}(k_{x})$ and $\mathbf{r}%
_{2}(k_{y})$. The linking number is a topological invariant that
characterizes the number of times that $\mathbf{r}_{1}(k_{x})$ and $\mathbf{r%
}_{2}(k_{y})$ wrap around each other \cite{Hannay05}. Thus, the spin-Chern
number is equivalent to the linking number of two closed curves $\mathbf{r}%
_{1}(k_{x})$ and $\mathbf{r}_{2}(k_{y})$ of the effective magnetic field $%
\mathbf{d}\left( \mathbf{k}\right) $ for the pseudospin-up Hamiltonian $%
h_{\uparrow }\left( \mathbf{k}\right) $. The spin-Chern numbers shown in
Fig.~\ref{figS2} are in accords with the representative links shown in
Fig.~2(a) of the main text.

For the pseudospin-down mode, the signs of the horizontal couplings $J_{1}$
and $J_{2}$ change into the opposite $-J_{1}$ and $-J_{2}$. The Hamiltonian
reads $h_{\downarrow }\left( \mathbf{k}\right) =-d_{0}\sigma _{0}+\left( 
\mathbf{d}_{x}\mathbf{,d}_{y}\mathbf{,}-\mathbf{d}_{z}\right) \mathbf{\cdot
\sigma }$. The spin-Chern number obtained from the linking of the effective
magnetic field $\left( \mathbf{d}_{x}\mathbf{,d}_{y}\mathbf{,}-\mathbf{d}%
_{z}\right) $ yields $C_{\downarrow }=-C_{\uparrow }$. This is
straightforward because that the curve $\mathbf{r}_{2}(k_{y})$ and two
components of the curve $\mathbf{r}_{1}(k_{x})$ in the $x$ and $y$
directions are unchanged, and the component of the curve $\mathbf{r}%
_{1}(k_{x})$ in the $z$ direction changes into the opposite.

\subsection{C. Robustness of the antihelical edge states}

\begin{figure*}[htbp]
\includegraphics[bb=10 5 595 190,width=18 cm]{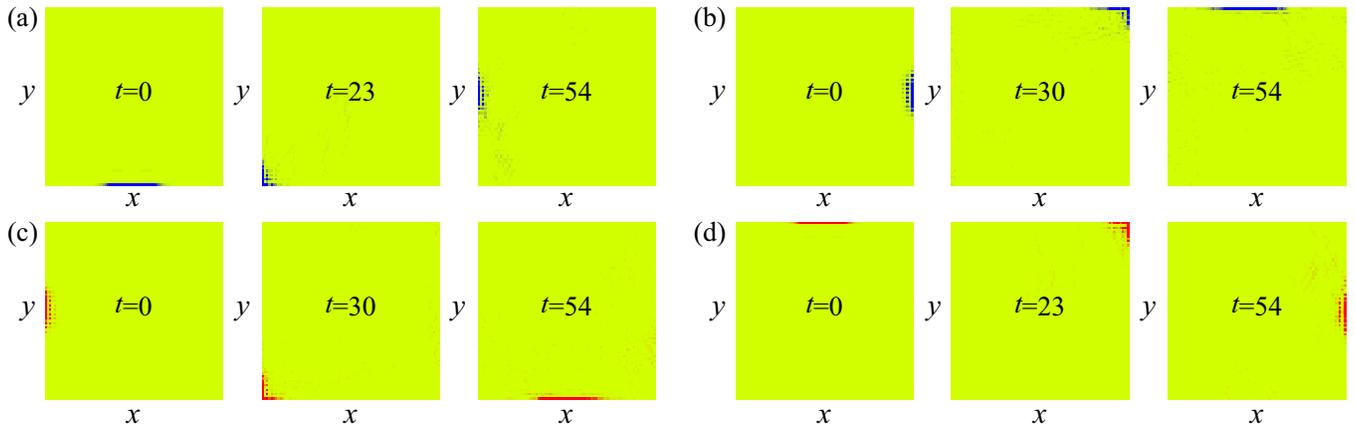}
\caption{Numerical simulations for the robust propagation in the square
lattice of size $80\times 80$. The parameters are $\protect\chi =\protect%
\kappa /2=1/2$, $J_{1}=J_{2}=1$, and $\protect\delta =3/2$, being identical
with the parameters in Fig.~\protect\ref{figS5}. The unit of time $t$ is $1/ 
\protect\kappa$.}
\label{figS6}
\end{figure*}

In the topological phases of the square lattice, both the helical edge
states and the antihelical edge states are topologically-protected and
robust to disorder. In comparison, the helical edge states in their
propagations are even more stable to the disorder because of the band gap
protection. In this section, we provide more details for the robustness of
antihelical edge states and their unidirectional propagations in the
presence of coupling disorder; notably, the couplings with random disorder
satisfy the particle-hole symmetry, which also exists in the square lattice
for the pseudospin-up (pseudospin-down). The numerical simulations of the
band structures and the robust propagations of antihelical edge states are
shown for the imperfect square lattice. The localized edge states and the
extended bulk states are spatially separated. Thus, the imperfection in the
bulk almost does not affect the edge states and their robust propagations.

The distribution and the localization of the edge states on the boundaries
are insensitive to the lattice size; however, the distribution and the
extended feature of the bulk states closely depend on the lattice size. The
larger lattice size leads to a better spatial separation between the edge
states and the bulk states. The influence on the edge states for the
disorder in the bulk of the square lattice is slight in comparison with the
influence on the edge states for the disorder on the boundaries of the
square lattice. This point is elucidated in Fig. \ref{figS4}, where we
depict the energy bands for the disorder in the bulk and on the boundaries,
respectively; and the random disorder is chosen on all the couplings either
inside or outside the central one-half area of the square lattice as
schematically illustrated in the upper panels of Fig. \ref{figS4}. The edge
states are depicted in Figs. \ref{figS4}(a) and \ref{figS4}(c) for the
random disorder in the couplings inside the central one-half area of the
square lattice in the bulk; and the edge states are depicted in Figs. \ref%
{figS4}(b) and \ref{figS4}(d) for the random disorder in the couplings
outside the central one-half area of the square lattice on the boundaries.
To obtain the 1D projection lattices, the 2D square lattice is set
translationally invariant in the $x$ ($y$) direction for Figs. \ref{figS4}%
(a) and \ref{figS4}(b) [Figs. \ref{figS4}(c) and \ref{figS4}(d)].

In the numerical simulations, the initial excitation has the Gaussian
profile in the form of%
\begin{equation}
\left\vert \Psi (t_{0})\right\rangle =\Omega
^{-1/2}\sum\nolimits_{k_{\varepsilon }}e^{-(k_{\epsilon
}-k_{0})^{2}/(2\alpha ^{2})}e^{-iN_{c}(k_{\epsilon }-k_{0})}\left\vert \psi
\right\rangle ,
\end{equation}%
where $\left\vert \psi \right\rangle $ is the edge mode with the momentum $%
k_{\epsilon }=k_{x}$ or $k_{y}$, and $k_{0}=\pi $. $N_{c}$ is the center of
the wave packet and $\alpha $ controls the width of the Gaussian profile. $%
\Omega $ normalizes $\left\vert \Psi (t_{0})\right\rangle $. The numerical
simulations of the robust light propagation for the excitations of
antihelical edge states in the presence of random disorder are performed and
demonstrated in Fig.~\ref{figS5}. All the couplings with random disorder are
deviated from the set parameters of the square lattice within the range of $%
[-10\%,10\%]$. The edge mode excitations robustly propagate along the
boundaries. The dynamics in the numerical simulations are close to the
dynamics in the absence of disorder exhibited in the main text Fig. 2; this
is a consequence of the limited lattice size $40\times 40$ in the numerical
simulations. The robust propagation against disorder is more excellent in
the numerical simulations performed in the larger size system as shown in
Fig.~\ref{figS6}.

In the topological metal phase possessing the antihelical edge states in
both the horizontal and vertical directions, there are two types of
right-angled ($90^{\circ }$ degrees) corners in the square lattice. One type
of corners help the right-angled turning of the edge mode excitations when
they propagate along the boundaries; and the other type of corners help the
edge mode excitations entering the bulk of the square lattice. Each corner
is associated with a cross coupling; and the two types of corners are
distinguished from their associated cross couplings $\chi +\delta $ and $%
\chi -\delta $. The formation of the right-angled corners associated with
the weak cross coupling enables the $90^{\circ }$ degrees turning across the
corners for the edge mode excitations propagating along the lattice
boundaries. The formation of the right-angled corners associated with the
strong cross coupling enables the edge mode excitations entering the bulk of
the square lattice.

\end{document}